# Thermal conductivity of a new carbon nanotube analogue: the diamond nanothread

Haifei Zhan[1,2,3], Gang Zhang[2,*], Yingyan Zhang[4], V. B.C. Tan[3], John M. Bell[1], and Yuantong Gu[1,*]

[1]School of Chemistry, Physics and Mechanical Engineering, Queensland University of Technology, Brisbane QLD 4001, Australia
[2]Institute of High Performance Computing, Agency for Science, Technology and Research, 1 Fusionopolis Way, Singapore 138632
[3]Department of Mechanical Engineering, National University of Singapore, 9 Engineering Drive 1, Singapore 117576
[4]School of Computing, Engineering and Mathematics, University of Western Sydney, Locked Bag 1797, Sydney NSW 2751, Australia

**Abstracts:** Based on the non-equilibrium molecular dynamics simulations, we have studied the thermal conductivities of a novel ultra-thin one-dimensional carbon nanomaterial - diamond nanothread (DNT). Unlike single-wall carbon nanotube (CNT), the existence of the Stone-Wales transformations in DNT endows it with richer thermal transport characteristics. There is a transition from wave-dominated to particle-dominated transport region, which depends on the length of poly-benzene rings. However, independent of the transport region, strong length dependence in thermal conductivity is observed in DNTs with different lengths of poly-benzene ring. The distinctive SW characteristic in DNT provides more degrees of freedom to tune the thermal conductivity not found in the homogeneous structure of CNT. Therefore, DNT is an ideal platform to investigate various thermal transport mechanisms at the nanoscale. Its high tunability raises the potential to design DNTs for different applications, such as thermal connection and temperature management.

*Corresponding author. E-mail: zhangg@ihpc.a-star.edu.sg (Gang Zhang); E-mail: yuantong.gu@qut.edu.au (Yuantong Gu)

## 1. Introduction

Carbon-based nanomaterials, such as zero-dimensional (0D) fullerene $C_{60}$, one-dimensional (1D) carbon nanotube (CNT), and two-dimensional (2D) graphene, have attracted great interests from the scientific and engineering communities since their first discovery [1, 2]. Take the 1D CNTs for example, their unprecedented properties have made them ideal building blocks for composite materials, thin films and nanoelectromechanical systems. Researchers reported that CNT-based mechanical resonators have a quality factor as high as 5 million [1]. Diverse commercial products incorporating bulk CNT powders ranging from rechargeable batteries, automotive parts and sporting goods have already emerged [3].

The fundamental properties of CNTs are determined by their diameter and charity. Many studies have shown that thin CNTs possess superior property over those with larger diameters. For instance, a sharp reduction in the elastic bending modulus (from 1 to 0.1 TPa) of CNTs is found when the CNT's diameter increases from 8 to 40 nm [4]. Also, ultra-thin CNTs are reported to show much higher carrier mobility [5] and exhibit superconductivity at temperature below 20 K [6]. The reported successful synthesis of ultra-thin CNTs including (2,2), (3,1) and (4,0) CNTs [7, 8]. Along with efforts to synthesize ultra-thin CNTs, a recent study reported the synthesis of a new ultra-thin CNT analogue (through solid-state reaction of benzene under high-pressure), termed as diamond nanothread (DNT) [9]. The DNT is a close-packed $sp^3$-bonded carbon structure, with carbon atoms arranged in a diamond-like tetrahedral motif (see Figure 1). According to the experimental and modeling works, the DNTs can be regarded as hydrogenated (3,0) CNTs connected with Stone-Wales (SW) transformation (see inset of Figure 1a) [10]. The distributed SW transformation interrupts the central hollow of the structure, differentiating DNTs from CNTs. Owing to their intrinsic structure, DNTs are envisioned to possess good load transfer rate through covalent bonding, thus allowing for technological exploitation in nanobundles or fabrics. A recent in silico studies show that the DNT has excellent mechanical properties [11], namely, stiffness of about 850 GPa, elongation to failure of 14.9%, and bending rigidity of about $5.35 \times 10^{-28}$ N·m$^2$.

Thermal transport in carbon nanostructures has been a research focus in past decade. Different experimental and theoretical works have been conducted to study the phonon transport mechanism in 1D carbon nanotubes and 2D graphene [12-18]. Length dependent thermal conductivity in CNT has been predicted by molecular dynamics simulations [19, 20] and then observed experimentally [21]. The physical connection between energy super-diffusion and thermal conduction is responsible to this length dependence [22]. It was demonstrated that in carbon nanotube, the thermal conductivity is dominated by phonons at all temperatures [23]. Kim *et al* measured the thermal conductivity of a single CNT and found that its room temperature thermal conductivity was larger than 3000 W/ mK [24]. Various effects on thermal conductivity of CNT, such as diameter [20], isotopic doping [20, 25], atom substitutional impurities [26], surface chemisorbed molecules [27], vacancies and Stone–Wales (SW) defect [28, 29], have been systematically explored. As a new type of 1D carbon nanomaterial, it is of great interest to know how the thermal transport properties of DNT differ from those of CNT. Thus, large-scale nonequilibrium molecular dynamics (NEMD) simulations were employed to probe the thermal transport properties of DNTs. In the next section, we introduce the computational methods and also DNT models adopted. This is followed by the determination of the thermal properties of the DNT, including their scaling characteristics, length dependence, and temperature effects. Finally, the key findings of this work are summarized.

2. **Computational Methods**

The DNT model was established based on recent experimental observations and first principle calculations [9]. Different DNT models were achieved by varying the number of poly-benzene rings between two adjacent Stone-Wales (SW) transformations. The C-C and C-H atomic interactions were modeled by the widely used adaptive intermolecular reactive empirical bond order (AIREBO) potential [30]. It has been shown to well represent the binding energy and elastic properties of carbon materials. The DNT models were firstly optimized by the conjugate gradient minimization method and then equilibrated using Nose-Hoover thermostat [31, 32] under ambient conditions for 4000 ps (i.e., at temperature of 300 K and pressure of 1 atm). Periodic boundary conditions were applied along the length direction during the relaxation process.

As illustrated in Figure 1a, the poly-benzene rings in the DNT segments are connected by SW transformations. A DNT unit cell with *n* poly-benzene rings between two adjacent SW transformations is denoted by DNT-*n*, e.g., DNT-8 has eight poly-benzene rings with a length approximating 4 nm (Figure 1a). Initial calculations show that the potential energy per carbon atom decreases sharply when *n* increases from 2 to 8, and converges to the value for the carbon atom in a hydrogenated (3,0) CNT (Figure 1b). Such results indicate that adding more SW transformations to the DNT structure of a given length would increase the total system potential energy, and thus reduce its stability. Therefore, following a previous study [33], the minimum number of the poly-benzene rings within in a DNT unit considered herein is 8 to ensure the stability of the structure.

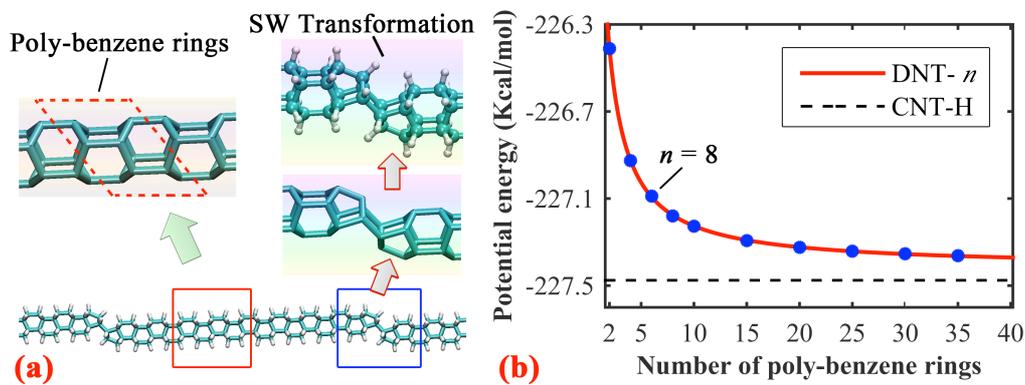

**Figure 1** (a) A segment of the DNT-8, insets show the structural representation of the poly-benzene rings and the Stone-Wales (SW) transformation. (b) The potential energy per carbon atom for different DNT units, CNT-H represents the hydrogenated (3,0) CNT.

Non-equilibrium molecular dynamics (NEMD) simulations were employed to calculate the thermal conductivity ($\kappa$) of the DNT at 300 K. After attaining the fully relaxed configuration, the model was firstly switched to non-periodic boundary condition. The two ends of the sample were fixed with two adjacent regions being set as the heat source and sink, respectively. The temperatures of the heat source and sink were kept at 310 and 290 K respectively by using the Langevin thermostat [34]. The system was firstly simulated for 20 ns to arrive a steady state, and continued to another 20 ns for the thermal conductivity calculation. To ensure reliable calculations, $\kappa$ is calculated at time intervals of 2.5 ns and averaged over 20 ns. All the MD simulations were performed using the software package *LAMMPS* [35] with a small time step of 0.5 fs.

3. **Results And Discussions**

*3.1 Thermal transport properties of DNT*

Figure 2a illustrates the temperature profile for DNT-55 (a single SW transformation in a sample length around 24 nm) obtained from NEMD simulations. Here temperature is defined as corresponding to the mean kinetic energy. Unlike CNTs, we observed a clear temperature jump ($\delta T$) at the region with SW transformation, signifying the existence of interfacial thermal resistance or Kapitza resistance (KR). The interfacial temperature jump is defined by extrapolation of the two linear fits to the interface. Different from the traditional circumstance that the KR is triggered when the heat flows across the contacting interface between different materials, the observed KR occurs between same poly-benzene rings. Due to the occurrence of such temperature drop, the temperature gradient along the heat flux direction ($z$) is approximated by $\Delta T / L$. Here, $\Delta T = T_h - T_c$ is the temperature difference between the heat source $T_h$ and heat sink $T_c$, and $L$ is the length of DNT. According to Fourier's law, the thermal conductivity ($\kappa$) can be estimated by $\kappa = -(Q / A\Delta t)/(\Delta T / \Delta z)$. Here, $Q / A\Delta t$ is the heat flux with $Q$ as the energy change, $A$ as the cross-sectional area and $\Delta t$ as the time interval. The whole DNT is approximated as a solid cylinder with a diameter of 0.5 nm (the approximate distance between exterior surface hydrogens), same as used previously by Roman *et al* [33].

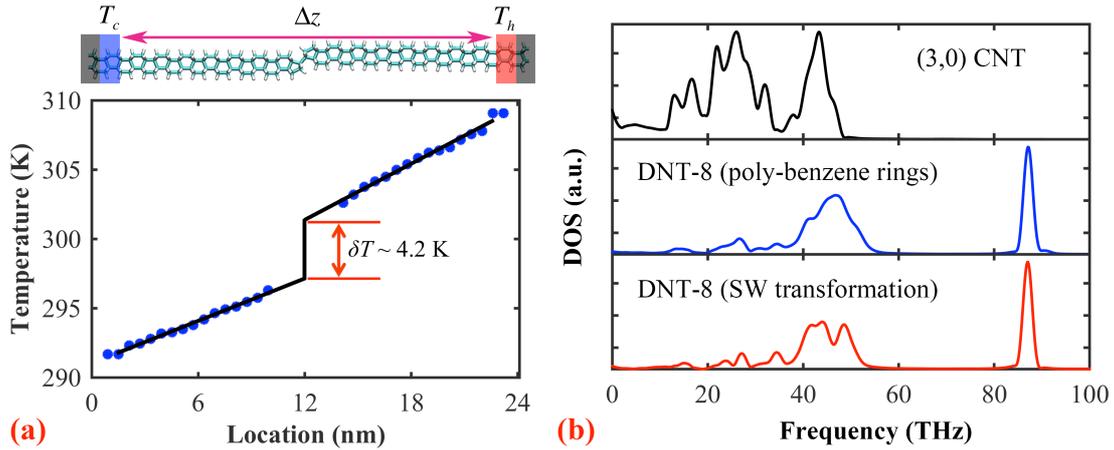

**Figure 2** (a) Temperature profile of the DNT-55 (length of ~ 24 nm), which contains only a single SW transformation. The atomic configuration is only a schematic representation of the studied model. (b) Comparisons of the DOS between CNT, the poly benzene rings portion of DNT, and the SW transformation region of DNT. The DOS is extracted from the autocorrelation function of the atomic velocities obtained from the MD simulations [36].

For comparison purpose, we also estimated the thermal conductivity of the (3,0) CNT by approximating its cross-sectional area as $\pi(D+h)^2/4$ (with $D$ and $h$ representing the diameter of the CNT and the graphite interlayer distance, respectively). Consistent with a recent study on ultra-thin (2,1) CNTs [37], the (3,0) CNT exhibits a low thermal conductivity of around 63.1 ± 6.3 W/mK. This is still about twice of that obtained from its DNT counterpart with a single SW transformation (~ 35.6 ± 4.7 W/mK). Acquiring the vibrational density of states (VDOS) of a (3,0) CNT, the poly-benzene rings and the SW transformation region of the DNT-8, a very different pattern was observed as shown in Figure 2b. Surprisingly, we found a great suppression of the phonon modes at frequencies below 60 THz in DNT (comparing with CNT), and the appearance of a new phonon peak at around 90 THz, which arises from the C-H bond vibrations. A similar additional phonon peak (~120 THz) is also reported by Zhang *et al* for graphene functionalized with –OH [38]. Such different DOS patterns are supposed to be responsible for the different estimated thermal conductivities between CNT and DNT. To unveil the different thermal transport properties between the (3,0) CNT and DNT-8, we have calculated the mode lifetimes using the recently developed approach by Daw *et al* [39, 40]. In Figure 3, we show the frequency dependent phonon lifetimes in pristine CNT and DNT-8, respectively. Compared with CNT, there is significant reduction in phonon lifetimes almost over the entire range of frequency in DNT-8. The reduction results from additional phonon scattering due to SW transformation, which explicitly explains the lower thermal conductivity in DNTs.

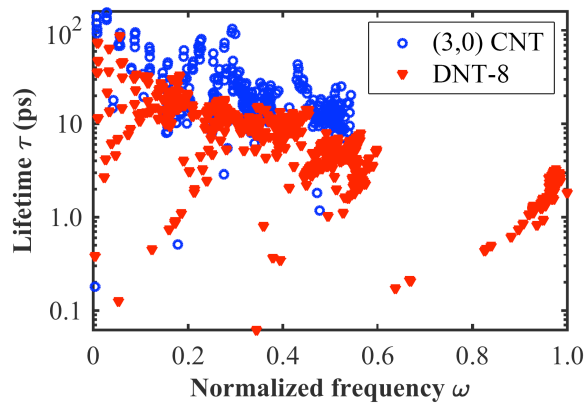

**Figure 3** The mode lifetimes as a function of the normalized frequency for (3,0) CNT and DNT-8.

Additionally, we should note that the DOS is largely determined by the employed potentials in MD simulation. For the studied system with $sp^3$ C-C bond, the recently developed ReaxFF potential [41] is also applicable. Our calculations reveal that the obtained radial distribution function (RDF) between ReaxFF and AIREBO potential are consistent with each other (Figure 4a), and a similar DOS pattern was obtained from ReaxFF potential (though the new phonon peak is around 100 THz, see Figure 4b). Considering that the thermal conductivities for most of CNT-related structures are derived from AIREBO potential so far, and the ReaxFF potential usually consumes huge computational resources, all our following simulations will use this potential. Also, we will emphasize the relative or normalized thermal conductivities of different DNTs, rather than their absolute values.

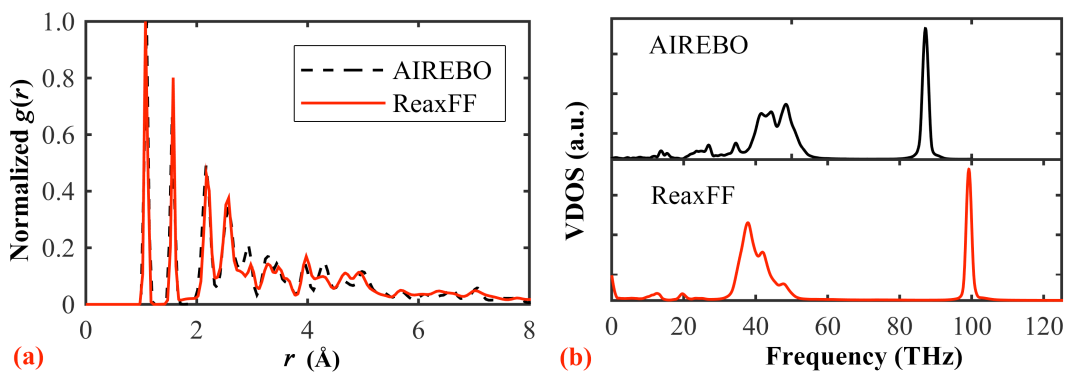

**Figure 4** Comparisons of (a) the RDF and (b) the DOS obtained from DNT-8 using AIREBO and ReaxFF potentials.

*3.2 Scaling behaviour*

The mismatch of the DOS (Figure 2b) between the section of the poly-benzene rings and SW transformation in the DNTs indicates phonon scattering at the SW region, which induces the abrupt temperature drop. To further explore this phenomenon, we compared the estimated thermal conductivities of DNTs with different sample lengths ($L$ = 24, 31, and 41 nm) and comprised by different constituent units (Figure 5a). Generally, the thermal conductivity increases as the sample length $L$ increases, irrespective of the number of poly-benzene rings. In particular, for a given sample length, as the number of poly-benzene rings increases, the estimated thermal conductivity initially decreases for smaller $n$, and then undergoes a relatively smooth increase. The most striking feature is the existence of a minimum in the thermal conductivity, which appears when $n$ is around 10, independent of the sample length.

Similar phenomena have also been observed in thin-film [42], nanowire [43] and superlattices [44, 45]. These have been explained from different perspectives.

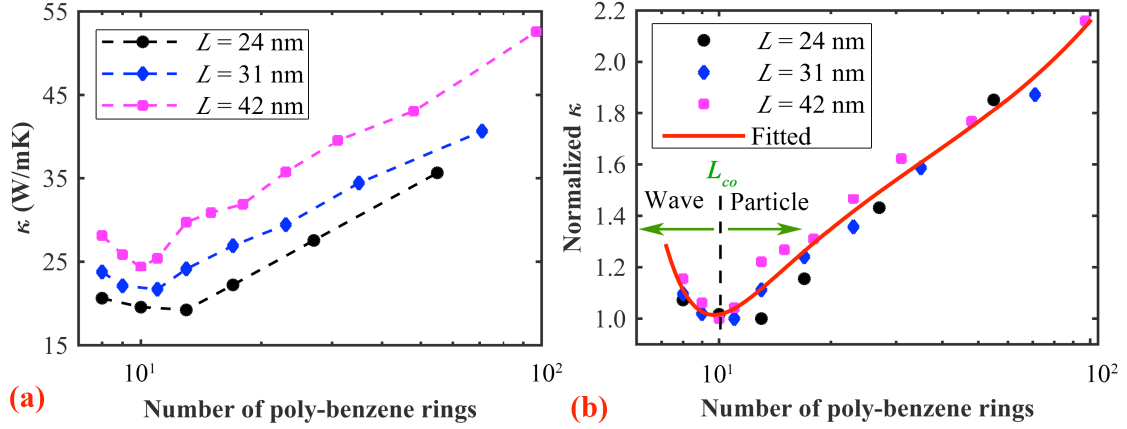

**Figure 5** (a) Thermal conductivity plotted as a function of DNT length and constituent unit cells; (b) The thermal conductivity normalized by its respective minima shows a general valley trend. This indicates universal critical length scales which correspond to the phonon coherence length $L_{co}$, estimated as 5.6 nm.

In line with the observations reported for graphene/h-BN superlattices [44], we obtain a single curve after normalizing the results by the minimum thermal conductivity, signifying a consistent scaling behaviour. Such behaviour can be explained from the perspective of the phonon coherence length as popularly adopted to analyse the thermal transport properties in superlattices. Generally, the minimum thermal conductivity occurs at a transition from wave-dominated to particle-dominated transport region [44]. That is, the phonon coherence length of the studied DNT corresponds to the length of a unit of DNT-$n$, with $n$ around 10 (see Figure 5b). In the left domain ($n < \sim 10$), phonon transport is largely dominated by wave effects including constructive and destructive interferences arising from interfacial modulation. While, in the right region, phonon waves lose their coherence and transport is more particle-like.

Further attempts have been made to estimate the phonon coherence length of the DNT based on the phonon DOS [44, 46], which is defined by $L_{co} = \tau_{co} v_D$ with $\tau_{co}$ and $v_D$ being the coherence time and Debye velocity, respectively. The coherence time is estimated according to $\tau_{co} = \int_0^{\omega_m} [g(\omega) D_{BE}(\omega)]^2 d\omega$. Here, $\omega_m$ is the maximum frequency of all phonon modes, and $g(\omega)$ is the normalized DOS according to the Bose-Einstein distribution $D_{BE}$ (i.e., $\int_0^{\omega_m} g(\omega) D_{BE}(\omega) d\omega = 1$). The Debye velocity is estimated from $3/v_D^3 = 1/v_L^3 + 2/v_T^3$, with $v_L$ and $v_T$ representing the sound velocity for the one longitudinal and two transverse branches, respectively [47]. Specifically, the sound velocity can be calculated from $v_L = \sqrt{(\lambda+2\mu)/\rho}$ and $v_T = \sqrt{\mu/\rho}$, respectively, based on the elastic Lame's constants $\lambda$, $\mu$, and the mass density $\rho$ [48]. However, a complete set of the mechanical properties of the fully hydrogenated single-wall CNT is unavailable, especially for the ultra-thin (3,0) CNT. Thus, we adopt the sound velocity for the single-wall carbon nanotube [49], i.e., $v_L = 1.3\times10^4$ and $v_T = 5\times10^3$ m/s, which leads to a Debye velocity around $5.67\times10^3$ m/s. The phonon coherence length is thus estimated for the DNT (with a length of ~ 42 nm) with different number of poly-benzene rings. As illustrated in Figure 6, a mean value of $L_{co}$ is around 5.6 nm, approximately the length of a DNT-11 unit cell. This appears to be in good agreement with the results presented in Figure 5b. We should note that such estimation is very primitive and has relatively large fluctuations due to the involved uncertainties (related to the adopted mechanical properties and DOS). However, this initial estimation and the consistent uniform scaling behavior of the thermal conductivity suggest that the existence of the SW transformations endows the DNT with a superlattice thermal transport characteristic.

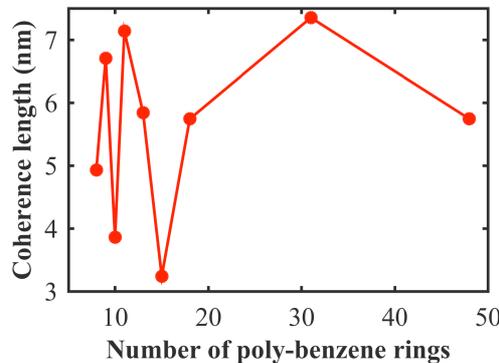

**Figure 6** The estimated phonon coherence length as a function of the number of poly-benzene rings for the DNT with a length around 42 nm.

*3.3 Length dependence*

The discussions above reveal that DNTs possess a superlattice characteristic that is different from CNT. Further studies were then dedicated to investigate the dependency of their thermal conductivity on DNT length. Two groups of DNTs were simulated, with one group as comprised by DNT-8 (lower than the coherence length) and another as comprised by DNT-23 (greater than the coherence length). Similar to the observations on the ultra-thin CNTs,[37] the thermal conductivities of both groups exhibit a strong length dependence. According to Figure 7a, the thermal conductivity increases evidently with the increase of length.

Moreover, according to Schelling *et al* [50], when the simulation cell size $L$ is not significantly longer than the phonon mean-free path (MFP), the thermal conductivity will be limited by the system size due to scattering at the interfaces with the heat source and sink. In such circumstances, the thermal conductivity can be approximated from $1/\kappa = 1/\kappa_\infty (1 + \lambda/L)$, according to the thermal conductivity in kinetic theory [51]. Here $\kappa$ and $\kappa_\infty$ are the size dependent and converged (when sample size is large enough) thermal conductivity, respectively. $L$ is the sample length, and $\lambda$ is the MFP. Such heuristic relationships indicate that the inverse of thermal conductivity is proportional to the inverse of the simulation cell length, and its intercept at the origin should be equal to the inverse of thermal conductivity of an infinitely large system. This linear scaling behavior of thermal conductivity has been commonly found in 1D nanostructures [51-54], 2D nanoribbons [55, 56], and bulk materials [50]. As illustrated in Figure 7b, the inverse of the estimated $\kappa$ for the DNT also follows a linear scaling with $1/L$. The fitting results suggest that $\kappa$ will saturate to 55.18 and 71.79 W/mK for the DNT-8 and DNT-23, respectively. Comparing with the MFP of single-wall CNT (~700-750 nm at room temperature [12]), the estimated MFPs for the two groups of DNTs (about 40.18 and 42.88 nm, respectively) are more than one order smaller, which is in consistent with the estimated lower thermal conductivity of the DNT.

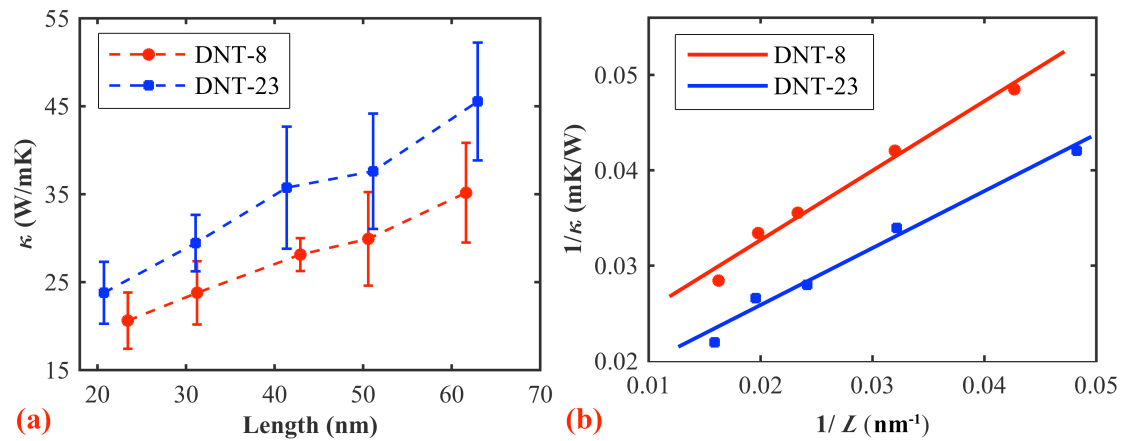

**Figure 7** (a) The thermal conductivity of DNT as a function of its length. (b) The inverse of the thermal conducted versus the inverse of sample length, which exhibits a common heuristic linear scaling relationship.

*3.4 Temperature dependence*

Finally, we assessed the thermal conductivity of DNTs at different temperatures. Figure 8 shows the thermal conductivity, which gradually decreases with an increase in temperature. It roughly obeys the $T^{-\alpha}$ law, but there is an obvious deviation from the normally predicted $T^{-1}$ relationship. In the Boltzmann transport equation, the phonon thermal conductivity is expressed as $\kappa(T) = \sum_j C_j(T) v_j^2 \tau_j(T)$ at a given temperature [57]. Here $C_j$, $v_j$ and $\tau_j$ are the specific heat, phonon group velocity and phonon relaxation time of phonon mode *j*. According to the Matthiessen rule [58], the phonon relaxation time is mainly limited by boundary scattering and the three-phonon umklapp scattering processes, i.e., $1/\tau = 1/\tau_b + 1/\tau_u$. Here $\tau_b$ and $\tau_u$ represent the relaxation-time parameters for boundary scattering and three-phonon umklapp scattering process, respectively. As the temperature increases, the higher-energy phonons are thermally populated which will enhance the role of umklapp scattering in determining $\kappa$ (i.e., $\tau_u$ will dominate the overall phonon relaxation time $\tau$). In Eucken's law, it states that the scattering rate of the Umklapp process at high temperature is proportional to temperature [59, 60]. As the specific heat is almost temperature independent, thus for homogeneous materials, their thermal conductivities drop steadily with the increasing temperature, and at high temperature, they obey a power-law dependence with an exponent of −1. However, in DNT, the additional phonon scattering at SW transformation weakens the role of umklapp scattering, supresses the temperature dependence. Therefore, the deviation from $T^{-1}$ law at high temperature is not surprising.

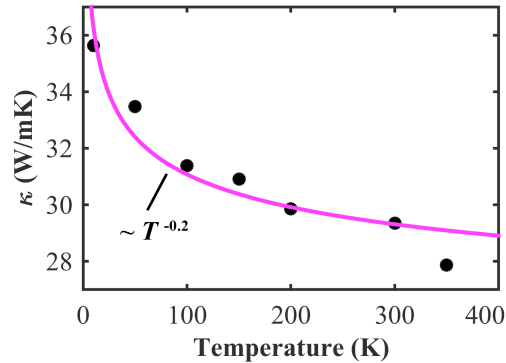

**Figure 8** Thermal conductivity of DNT-13 (size of ~ 42 nm) as a function of temperature. The solid line is fitted with a power-law relation.

## 4. Conclusions

In summary, we have studied the thermal conductivities of a novel ultra-thin one-dimensional carbon nanomaterial, namely, diamond nanothread (DNT), based on the non-equilibrium molecular dynamics simulations. Distinct from single-wall CNT, the DNT exhibits a relatively low and a strong length dependent thermal conductivity. Due to the existence of the Stone-Wales transformations, DNTs exhibit a superlattice thermal transport characteristic. That is the thermal conductivity shows a uniformly decreasing and then increasing profile with respect to the number of poly-benzene rings. Such phenomenon is supposed as resulted from the transition from wave-dominated to particle-dominated transport region. These findings have suggested a highly designable thermal transport property of DNT that is suitable for the thermal connection applications. As a thermal connection, in addition to a high absolute value of thermal conductivity, it is also expected to have a highly tunable thermal conductivity in order to match the working devices by certain selection rules. Further studies show that the thermal conductivity of DNT decreases continuously with increasing temperature.

**Acknowledgement**

Support from the ARC Discovery Project (DP130102120), the Australian Endeavour Research Fellowship, and the High Performance Computer resources provided by the Queensland University of Technology, University of Queensland, A*STAR Computational Resource Centre (Singapore), and University of Western Sydney are gratefully acknowledged.